\documentclass[reprint,showpacs,preprintnumbers,nofootinbib,amsmath,amssymb,aps,prd,floatfix,superscriptaddress,onecolumn,notitlepage]{revtex4-1}
\pdfoutput=1
\usepackage{graphicx}
\usepackage{bm}
\usepackage{subfigure}
\usepackage{url}
\usepackage[hyperindex]{hyperref}
\usepackage{color}
\usepackage[ddmmyy,24hr]{datetime}
\usepackage{bigdelim}
\usepackage{booktabs}
\usepackage{dcolumn}
\usepackage{multirow}
\usepackage{subfigure}
\usepackage{cancel}
\usepackage{stackrel}
\usepackage{paralist}
\usepackage{xspace}
\usepackage{slashed}
\usepackage{cancel}
\usepackage{todonotes}
\usepackage{ulem}

\newcommand{\nua}[1]{\ensuremath{\rlap{\kern-2.5pt\ensuremath{\overset{\scriptscriptstyle(-)}{\phantom{\nu}}}}{\ensuremath{{\nu}_{#1}}}}}

\usepackage{enumitem}

\makeatletter
\def\namedlabel#1#2{\begingroup
    #2%
    \def\@currentlabel{#2}%
    \phantomsection\label{#1}\endgroup
}
\makeatother
\begin{document}

\title{No CC-NSI explanation of the Gallium anomaly}

\author{C. Giunti}
\email{carlo.giunti@to.infn.it}
\affiliation{Istituto Nazionale di Fisica Nucleare (INFN), Sezione di Torino, Via P. Giuria 1, I--10125 Torino, Italy}

\author{C.A. Ternes}
\email{ternes@to.infn.it}
\affiliation{Istituto Nazionale di Fisica Nucleare (INFN), Sezione di Torino, Via P. Giuria 1, I--10125 Torino, Italy}


\begin{abstract}
We show that the Gallium anomaly can not be explained by CC-NSI.
\end{abstract}

\maketitle

In the first version of this article, we have argued that the Gallium anomaly can be resolved with charged current neutrino nonstandard interactions (CC-NSI). Unfortunately, it was not taken into account that
in the Gallium experiments (GALLEX, SAGE, and BEST) the neutrino flux was determined from the activity of the source, which was obtained from the measurement of the heat emitted by the source. Therefore, if CC-NSI were present in the production of the neutrinos, they would be already included in the calculation of the neutrino flux. Likewise, the cross-section
of the detection process $\nu_e + {}^{71}\text{Ga} \to e^- + {}^{71}\text{Ge}$
is calculated from the measured lifetime of ${}^{71}\text{Ge}$,
which determines the contribution of the transition to the ground state of ${}^{71}\text{Ge}$
(see, e.g., Eqs.~(3) and~(5) of Ref.~\cite{Semenov:2020xea}).
The contributions of the transitions to the excited states are calculated
trough the relative Gamow-Teller strength,
as shown, e.g., in Eq.~(6) of Ref.~\cite{Kostensalo:2019vmv}.
Hence,
if CC-NSI were present in the detection process,
they would contribute also to the lifetime of ${}^{71}\text{Ge}$
and
they would have already been included in the cross section of the detection process.
It can be concluded that if CC-NSI exist, they are included in
the standard estimation of the deficit observed in the Gallium experiments.

In this case,
the neutrinos produced and detected in the Gallium experiments
are described by the same \emph{normalized} state
\begin{equation}
|\nu_s\rangle
=
|\nu_d\rangle
=
|\nu\rangle
=
\dfrac{1}{N}
\left(
|\nu_{e}\rangle + \sum_{\beta=e,\mu,\tau} \epsilon_{\beta} |\nu_\beta\rangle
\right)
,
\label{state}
\end{equation}
where
\begin{equation}
N
=
\left(
1 + 2 \text{Re}[\epsilon_{e}]
+
\sum_{\beta=e,\mu,\tau} |\epsilon_{\beta}|^2
\right)^{1/2}
,
\label{N}
\end{equation}
such that
\begin{equation}
\langle\nu|\nu\rangle = 1
.
\label{normalization}
\end{equation}
Therefore,
the effective zero-distance survival probability
for the short baselines of the Gallium experiments,
where there are no oscillations
due to the standard solar and atmospheric squared-mass differences,
is equal to unity: 
\begin{equation}
P_{ee}^{\text{SBL}} = |\langle \nu_d | \nu_s \rangle|^2
=
|\langle \nu | \nu \rangle|^2 = 1
.
\label{Pee}
\end{equation}
We conclude that
the Gallium anomaly cannot be explained by CC-NSI.

\begin{acknowledgments}
We thank Mariam T\'ortola and David Vanegas Forero for discussions on this topic.
\end{acknowledgments}


%

\end{document}